# Distribution-theoretic methods in quantum field theory[1]

F.V. Tkachov

*Institute for Nuclear Research of Russian Academy of Sciences*
*Moscow, 117312, Russian Federation*

The evolution of the distribution-theoretic methods in perturbative quantum field theory is reviewed starting from Bogolyubov's pioneering 1952 work with emphasis on the theory and calculations of perturbation theory integrals.

I would like to review the origin and evolution of the idea that generalized functions (distributions) play with respect to integrals of perturbative quantum field theory (pQFT) a role similar to that of complex numbers with respect to polynomials.

What makes this topic interesting and instructive is the apparent contradiction between the proven power of the distribution-theoretic methods in pQFT (it would be enough to mention the Bogolyubov $R$-operation) and the low awareness of theorists of this powerful technique.

SOME HISTORY. Singular generalized functions had been emerging in various applications (recall Dirac's $\delta$-function). Sobolev [1] found a systematic way to define such generalized functions as linear functionals on suitably chosen spaces of test functions. The most universally useful variant of the theory was suggested by Schwartz [2] (the so-called distributions; I will use the term interchangeably with generalized functions) who made a great effort to propagandize the simplicity and power of the technique of distributions [3] as did Gelfand and Shilov [4]. There is nothing inherently difficult in the idea of generalized function, and the theory can be taught in an almost elementary fashion [5]. Distributions ought to be taught to students early because of their wide usefulness (e.g. [3], [4], [6]).

As a side remark, I'd like to point out that the generalized functions defined as linear functionals on test functions are much more sensible candidates for the role of "arbitrary functions" than the usual interpretation in terms of an arbitrary correspondence between arguments and values. Any constructive mathematical object must exist in the form of approximations expressible via finite symbolic sequences. This holds true for continuous functions (approximations via linear splines with rational nodes), and for generalized functions defined as linear functionals — but not for the conventional "arbitrary" functions. Furthermore, if one obtains the value of a function's argument from some measurements involving say statistical errors, then all one can directly measure is an average of the function values — which is immediately interpreted as an integral of the function against a special test function that describes the distribution of errors. Thus generalized functions as linear functionals directly correspond to the reality of physical measurements of functional dependences.

From [7] one learns that Bogolyubov's teacher and collaborator of many years N.M. Krylov was interested in generalized solutions of differential equations, so a reasonable guess is that Bogolyubov and Krylov studied Sobolev's ideas soon upon publication. As a result, Bogolyubov was heuristically prepared to accomplish the conceptual breakthrough in regard of the problem of UV divergences in the early 50's [8] — but was not yet familiar with the smooth technique of distributions [2]–[5] to make a systematic use of it in formal proofs.

In a remarkable letter [8], Bogolyubov pointed out that (i) UV divergences result from an incorrect formal treatment of products of singular functions, (ii) such products are not defined by physical principles at the points where the singularities overlap, and (iii) a correct way to define the amplitudes at such singular points is via the procedure known as extension of functionals (which is a basic tool of functional analysis, the simplest variant being the Hahn-Banach theorem). So, ref. [8] shed a scientific light on the problem of UV renormalization and reduced it to a more or less straightforward working out of the formulas and proofs for what became known as the $R$-operation [9]. However, despite the distinctly functional-analytic flavor of the reasoning in [8], the formal proofs in [9] were completely within the limits of the ordinary integral calculus. This is explained by the fact that a technique for handling multidimensional distributions was not available to the authors of [9].

Anyhow, the monograph [10] summarized Bogolyubov's findings, and the theoretical community was presented with both a heuristic distribution-theoretic derivation of the $R$-operation, and a formal proof of the finiteness of the result in terms of the conventional integral calculus. The sad fact is, the derivation was almost universally ignored even by mathematical physicists who ought to know better. An exception was the work of Epstein and Glaser [11] who attempted to formalize Bogolyubov's construction in a distribution-theoretic manner, but the essential mechanism which trivializes the formula and the finiteness of the $R$-operation was not clarified and remained buried in many details pertaining to the operator specifics of $S$-matrix, etc. Ref. [11] remained largely unknown to the theoretical community.

On the other hand, the pro forma proof [9] (of no heuristic value whatsoever) received considerable attention. It was cor-rected by Hepp [12], and improved via the forest formula [13] (rediscovered in [14]). The works [9], [12]–[14] created what became known as the BPHZ theory. It is commonly treated as *the* theory of UV renormalization, which identification — I emphasize — is incorrect because the BPHZ formalism ignores the key ingredient of the discovery of *R*-operation, namely, the distribution-theoretic argument of [8].

---

[1] Talk at the Bogolyubov Conference on Problems of Theoretical and Mathematical Physics, Moscow-Dubna-Kyiv, 27 Sept. – 6 Oct. 1999.

The heuristic arguments used to obtain breakthrough results such as the *R*-operation are unquestionably more important than any artificial proof, and mathematical physicists ought to study and clarify such arguments in the first place. But this did not happen in reality.

DIGRESSION ON SIMILARITY OF UV RENORMALIZATION TO DIFFERENTIATION. A widely-spread attitude is that UV renormalization is an artificial procedure on top of ordinary integrals, a temporary prop, and it will go as soon as physicists find a better formalism to describe the domain of very high energies. In particular, the limit $\Lambda \to \infty$ (where $\Lambda$ is the intermediate UV cutoff) is considered unphysical and in need of eventual modification.

However, such an attitude has only psychological roots in an inadequate mathematical education. Indeed, there is no principle which would restrict Nature in Her choice of mathematical objects in terms of which to formulate Her laws. In fact, the tremendous success of QED shouts at us to accept the objects of the type "integral + subtractions" as a whole — i.e. as hybrid objects which possess features of both ordinary integrals and generalized functions.

Furthermore, consider time derivatives in classical mechanics: $\dot{x} = \lim_{\Delta t \to 0}(\Delta x / \Delta t)$. Is not the limit $\Delta t \to 0$ as unphysical as $\Lambda \to \infty$? Is not the time derivative only a mathematical trick to allow a precise description of the Solar System? Would not physics need to be modified at small time scales? It does get modified as quantum effects come into play. However, not only the time derivatives are not eliminated but quantum mechanics introduces spatial derivatives. And the transition to still smaller space-time scales in relativistic quantum theory introduces UV renormalization which, although not exactly a differentiation, can be regarded as an operation of the same general type: $RG = \lim_{\Lambda \to 0} Z_\Lambda G^\Lambda$. To summarize:

| | |
|---|---|
| Classical mechanics | $\partial/\partial t$ |
| Quantum mechanics | $\partial/\partial t$, $\partial^2/\partial x^2$ |
| Quantum field theory | $\partial/\partial t$, $\partial^2/\partial x^2$, $R$ |
| Quantum gravity | $\partial/\partial t$, $\partial^2/\partial x^2$, $R$, ... |

The reality is, the deeper we go, the more numerous and various singular operations we encounter. From this viewpoint, the popular "super" theories that attempt to eliminate UV renormalization altogether are as bizarre as would be an attempt to eliminate time derivatives from the classical mechanics (imagine a theory of the Solar System with a superpartner for each planet, ... etc.)

In short, the UV renormalization is the sweeping of dust under the carpet to no greater extent than the use of time derivatives in classical mechanics.

DISTRIBUTIONS TRIVIALIZE THE BOGOLYUBOV-PARASYUK THEOREM. I already mentioned ref. [11] which attempted a mathematical clarification of Bogolyubov's construction of pQFT together with the *R*-operation. However, it was limited in scope and addressed a rather specific problem. In a more systematic manner the problem had to be addressed in the context of the theory of asymptotic operation (AO) to which I'll turn below (a review of the first, Euclidean part of the theory is given in [15]).

The background was as follows. I ran across the textbook [3] by chance in my second semester at the Moscow State University (1974), and read it because it is a highly readable collection of specific examples with the irrelevant abstract parts of the theory of distributions omitted. Now it seems to me that the "true" mathematical physicists who later pounced me on the head with their anonymous reports, learn distributions from the "real" book [2] which is entirely devoted to the abstract theory without a single meaningful example. Anyhow, I had a unique opportunity to go through the rest of my curriculum with a working knowledge of distributions, which proved to be highly beneficial, and I agree with Richtmeier [6] on how distributions should be taught (if what I learned from [3] were not useful I'd forgot it long ago as I did some other books I read then). By the time I took the course of quantum field theory (1975) I had distributions in my bones and had no trouble grasping the meaning of local UV counterterms and proceeded to develop a scenario for a distribution-theoretic proof of the *R*-operation as a mnemonic tool which allowed me to ignore the extremely cumbersome and unilluminating BPHZ arguments as they indeed deserve. The scenario was formalized in [16]. Unfortunately, I learned about ref. [11] much later and so was unable to employ the authority of Epstein and Glaser as a (much needed) protection against inconscientious referees from the BPHZ camp (one of whom — a very leading expert, evidently — proclaimed with a superb arrogance that "*the operator product expansion has been completely clear for more than 10 years. A new theory is not needed.*").

The techniques of distributions properly extended to many dimensions allows one to make a full use of the recursion structure of the problem and perform proofs inductively with respect to dimensionality of the manifolds on which singularities are localized. A discussion of this key dilemma (singularities *vs.* the recursion structure) is given in [15]; briefly speaking, the BPHZ method sacrifices the recursion in order to avoid distributions (and ends up as a result with the combinatorial monstrosity of the forest formula) whereas AO develops a technique to handle singularities in order to take full advantage of the recursion structure.

The importance of such a proof (which essentially trivializes the mechanism of finiteness of *R*-operation) is that — unlike the BPHZ-style proofs — it is directly coupled to the heuristics of the problem. I don't have space to explain technical details and refer to the review [15] where an example is given, and to a systematic formal exposition of [17].

One could argue that all one needs for practical purposes is a rule for subtractions that works, and whether or not the proofs are transparent is irrelevant. Unfortunately, life is not that simple because there are far too many problems where one needs to handle specific patterns of singularities (soft, collinear, mass shell…), and it is impossible to enumerate and memorize all possible cases and list all the corresponding rules. A more sensible approach is to have *a systematic rule to generate rules for doing subtractions in specific situations*. To ensure that results are correct the method must so immediately translate into formal proofs as to obviate them altogether.

From the distribution-theoretic viewpoint all such problems follow the same general pattern: singularities are generated by zeros of denominators; intersections of the corresponding singular manifolds require special treatment; they are enumerated in a straightforward fashion [the subgraphs]; the added counterterms are localized on those manifolds [such a manifold is described by a system of equations $P_i = 0$ where $P_i$ are the corresponding denominators, and the counterterms are simply products proportional to derivatives of $\Pi_i \delta(P_i)$].

ASYMPTOTIC OPERATION (AO). Starting from 1978 I was involved in pQCD calculations and, again, when confronted with the so-called mass singularities was quickly able to see (in the early 1981) the analogy with Bogolyubov's 1952 argument: a formal manipulation (in this case, a mass expansion) results in an infinite expression (a mass singularity) to correct which one adds a counterterm localized at the point of singularity, roughly like this:

$$\frac{1}{p^2 - m^2} = \frac{1}{p^2} + \frac{m^2}{p^4} + c(m)\delta(p) + \ldots$$

The new element here compared with UV divergences is the requirement of asymptotic smallness of the remainder, which, together with a key requirement that the resulting expansion runs in pure powers and logarithms of the expansion parameter [18], allows one to fix the finite part of $c(m)$. Coupled with the techniques for handling products of singular factors, this led to:

(i) theoretical results that the BPHZ experts had for years been struggling to obtain (short-distance expansion valid for models with massless particles [18]);

(ii) powerful calculational formulas that helped define state of the art in the field [18], [19];

(iii) extensions (including calculational formulas) to the entire class of expansions of Euclidean type [21] (including mass expansions).

The formulas obtained using the Euclidean AO in 1984–1986 formed a basis for a calculational industry (a number of such calculations have been used e.g. in the precision measurements at LEP1). Of course, they were rewritten in the style of BPHZ and subsequent references are made to those secondary publications. The objective reason was that in Euclidean problems such as OPE and mass expansions, singular distributions appear only in intermediate formulas and can be integrated out from the final answers. (There were also subjective reasons such as the lack of any formal mechanism in the international research community to protect creative individuals from plagiarism by representatives of well-established mafias. Watch [22] for more on this.)

NON-EUCLIDEAN EXTENSION OF AO. By 1990, with all the results of the Euclidean AO neatly appropriated by other authors (my senior colleagues; see [23] for some related bibliographic comments), I had no choice but to push forward in the non-Euclidean direction. Already in 1984 [20] I clearly understood both the fact that the then obtained formulas were limited to Euclidean problems[2], as well as the fact that the scheme of AO per se was in no way limited to Euclidean situations, and that the true challenge was the general asymptotic expansion problem for PT integrals in Minkowski (non-Euclidean) space.

However, both my collaborators and plagiarists poohpoohed the idea. This did not prevent me from participating in QCD calculations [24] with a view to extend the method of AO to pQCD problems of non-Euclidean type. I realized that the key difference from the Euclidean formulas is a secondary expansion for the counterterms (the homogenization) needed to achieve pure power-and-log dependence on the expansion parameters [25]. The last step [26] was to realize that if one follows the routine of AO in a systematic fashion, no special rules for correct scalings etc. are needed to do the power counting at the singularities localized on non-linear manifolds: All one has to do is perform the secondary expansion (the homogenization) in the sense of distributions (as is indeed warranted by the logic of the problem) — with all the corresponding counterterms, etc. In the language of AO, the difficulties which, say, pQCD experts encountered with power counting at mixed softcollinear singularities are due to the fact that the secondary expansions, if done formally, may result in non-integrable singularities to which the routine of AO has to be applied with appropriate (straightforward) modifications — but that is hard to see in the context of the forest formula.

The result of [26] represents the much needed "rule to generate rules" for doing expansions in specific non-Euclidean situations.

Needless to say that the truly huge physical importance of the problem of non-Euclidean expansions and a huge variety of specific problems (with the resulting poor communication between experts from different theoretical niches) creates an insurmountable temptation to devise a "method of regions" in order to rewrite the prescriptions obtained via AO in the BPHZ style — exactly as was done with the results of Euclidean AO.

Fortunately, the method of non-Euclidean AO is applicable to problems involving phase-space $\delta$-functions (because for AO such $\delta$-functions are not really different from ordinary propagators). So distributions cannot be eliminated from final answers thus defeating attempts to rewrite them BPHZ-style. An example of such a problem is the systematic perturbation theory with unstable fundamental fields described in [27].

There is another line of research that leads to distributions from another direction: the so-called algebraic (a.k.a. integration-by-parts) algorithms invented in [28] have by now become an indispensable tool for automated large-scale calculations of multiloop integrals [29]. A recent extension to loop integrals with arbitrary topologies and mass patterns [30] relies on results of the theory of singularities of differentiable mappings [31], and there is a connection with the homogenization procedure of the non-Euclidean AO [26]. But I've run out of space. Watch [22] for more.

This work was supported in part by the Russian Foundation for Basic Research under grant 99-02-18365.

---

[2] Not that there is a shortage of physically interesting problems of this kind. In fact, some my former colleagues seem to have chosen to forever remain experts in that kind of calculations.

hep-th/9911236